# Downtime-Aware O-RAN VNF Deployment Strategy for Optimized Self-Healing in the O-Cloud

Ibrahim Tamim
ECE Department
Western University
London, Canada
itamim@uwo.ca

Anas Saci
ECE Department
Western University
London, Canada
asaci@uwo.ca

Manar Jammal
School of IT
York University
Toronto, Canada
mjammal@yorku.ca

Abdallah Shami
ECE Department
Western University
London, Canada
abdallah.shami@uwo.ca

*Abstract*—Due to the huge surge in the traffic of IoT devices and applications, mobile networks require a new paradigm shift to handle such demand roll out. With the 5G economics, those networks should provide virtualized multi-vendor and intelligent systems that can scale and efficiently optimize the investment of the underlying infrastructure. Therefore, the market stakeholders have proposed the Open Radio Access Network (O-RAN) as one of the solutions to improve the network performance, agility, and time-to-market of new applications. O-RAN harnesses the power of artificial intelligence, cloud computing, and new network technologies (NFV and SDN) to allow operators to manage their infrastructure in a cost-efficient manner. Therefore, it is necessary to address the O-RAN performance and availability challenges autonomously while maintaining the quality of service. In this work, we propose an optimized deployment strategy for the virtualized O-RAN units in the O-Cloud to minimize the network's outage while complying with the performance and operational requirements. The model's evaluation provides an optimal deployment strategy that maximizes the network's overall availability and adheres to the O-RAN-specific requirements.

## I. INTRODUCTION

The need for mobile networks will increase as we move towards a more connected world. The number of connected devices is set to surpass three times the global human population by 2022 [1]. To deal with such large demands, Network Service Providers (NSPs) must offer a diverse set of services to cope with the expanding varieties of connected devices and their applications such as Internet of Things (IoT), Vehicle-to-Everything (V2X) communications, and extreme real-time communications. With 5G networks, NSPs can mitigate many diversification, latency, and scale challenges through the use of network slicing, Software Defined Networking (SDN), and Network Function Virtualization (NFV) technologies [2]. With the current age of next-generation mobile networks, three categories of next-gen services can be achieved, namely, Enhanced Mobile Broadband - eMBB (up to 10 Gbit/s), Ultra-Reliable and Low-Latency Communications - uRLLC (up to ~ 1 ms), and Massive Machine Type Communications (mMTC). With the emergence of 5G networks, users expect high Quality of Experience (QoE) with seamless services that are available anywhere anytime. However, millions of users can be disrupted

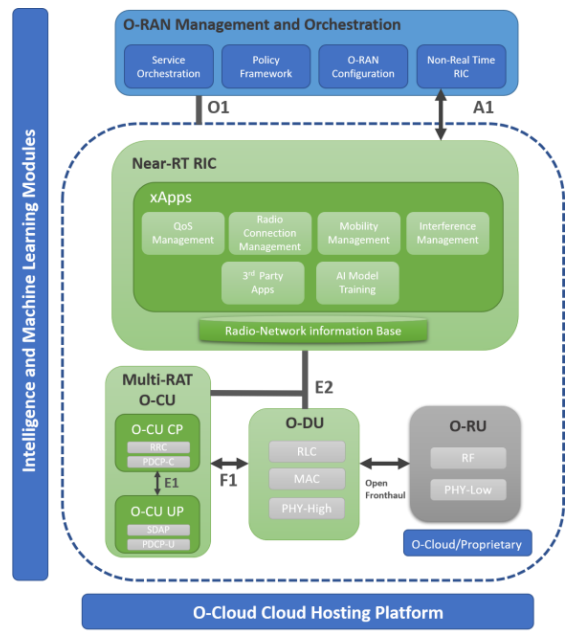

**Fig. 1**. The O-RAN architecture showing the implementation of intelligence modules across all layers, the O-Cloud platform, and within the dotted line, the O-RAN VNFs

due to network outages [3]. It is true that softwarization (including the virtualized cloudification) of the RAN units promises many performance-aware advantages, but its resiliency and availability are still key issues that should be addressed. 5G has low latency requirements; thus, manual outage management is not enough anymore. Therefore, absence of the proper outage management and compensation approaches does not only affect the repair process, but it defeats the 5G system purpose. Additionally, network growth and complexity put additional stress on the network operators' expenses, which are already very significant. Studies show that the node failure probability can reach 60-99% with the increase in the network density and radio nodes in 5G networks [4]. That said, these challenges can be greatly mitigated within the mobile network using SDN, NFV, and mobile edge computing (MEC) technologies [5]. These technologies can be easily integrated with the O-RAN due to its agility and openness [6]. In this case, the O-

RAN units can be hosted on the cloud to support dynamic service function chaining (SFC), network slicing, and dynamic scaling. Fig. 1 shows the O-RAN architecture. The standards and approaches of the O-RAN that can be used to enable cloudification and self-organization functionalities are still being researched and studied.

We aim to contribute to these developments by providing optimized self-healing functionalities for deploying O-RAN units. However, deploying these units is challenging as their location can greatly affect the network's availability, latency, cost, and other performance metrics. This work is the first step towards a full-scale optimized self-healing engine for O-RAN. In this paper, we present an optimization model to deploy the O-RAN units and their components (redundant and dependent) within the regional and edge clouds while minimizing the outage per unit and per SFC. In addition, we aim to provide fast recovery in case of failure in the units or their hosts. We summarize our main contributions as follows.

- Identify the abstract details of the O-RAN units and map them to the NFV infrastructure.
- Address O-RAN units' self-healing from an outage management approach to maintain availability baseline.
- Propose a resiliency-aware deployment strategy for the O-RAN units that integrates the performance (latency/computational) and availability constraints.
- Capture various availability aspects for the deployment approach, including redundancy models for the units, dependency relation between different units, and outage-related metrics of the nodes (failure rates and recovery times for servers and O-RAN units).
- Design an optimization model as the first building block of an intelligent approach for O-RAN self-healing.

The remainder of this paper is organized as follows; Section II covers the related work, Section III discusses the problem overview and modelling. Section IV covers the evaluation and use case, Section V presents the results, Section VI concludes the work, and Section VII presents our acknowledgment.

## II. RELATED WORK

Sharma et *al.* [8] design a provider network to achieve high-availability SFCs using disparate network components and low-availability Virtualized Network Functions (VNFs), however, they discard the dependency between different VNFs. Fan et *al.* [9] propose an online greedy algorithm to minimize physical resource consumption while meeting the client's SFC availability constraints and considering off-site redundant VNF components. Araujo et *al.* [10] focus on the decision of assigning backup SFC to fulfill its availability constraints while improving the resource efficiency across all VNFs. Jammal et *al.* [11] propose an optimization model for a regional cloud to enhance the availability of its applications while considering the multi-tier components' relationships. However, it discards low latency applications and edge-aware services and can only be applicable to applications of one regional cloud. These papers highlight the importance of availability in terms of cost, reliability, and quality of the network's services. Considering these factors and to the best of our knowledge, there has not been a proposed benchmark solution that addresses availability for the O-RAN cloudification use-case. In this paper, we treat availability as the main objective to achieve when hosting all the VNF components (O-RAN units) in the cloud while considering the specific O-RAN constraints such as latency between the VNF components, regional and edge placement constraints, and redundancy constraints. We provide an exact method of a binary integer programming (BIP) optimization model as a benchmark for optimizing self-healing in the O-RAN use case.

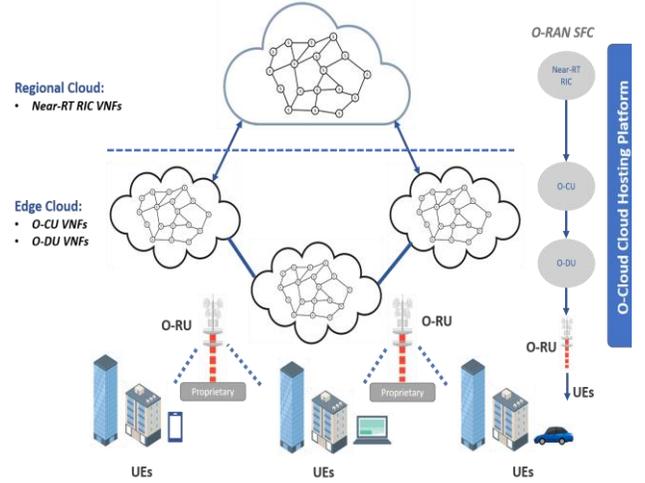

**Fig. 2**. O-RAN deployment scenario B showing the O-RAN SFC from the near-RT RIC in the regional cloud towards the UEs

## III. PROBLEM OVERVIEW AND MODELLING

With O-RAN, different vendors can be used to avoid single point of failure and vendor lock-in. However, such an advantage can add other limitations on the network. Adopting the concept of multiple vendors does not guarantee more reliable or secure solution than a proprietary one. When the O-RAN units are implemented on different software and hardware, new unexpected failures or vulnerabilities might emerge. With this complicated system and to maintain the 5G and O-RAN interoperability promises, it is important to propose a proactive outage-aware approach for managing those O-RAN units.

The O-RAN architecture consists of a Non-Real-time RAN Intelligent Controller (non-RT RIC) that provides machine learning model-training, data acquisition, service management, and policy generation; a near-RT RIC that hosts the micro-service-based applications controlling the RAN infrastructures; an O-RAN Central Unit (O-CU) that controls the radio protocol stacks, an O-RAN Distributed Unit (O-DU) that manage the physical layer functionalities, and lastly a Radio Unit (O-RU) that provides RF processing. In this architecture, the RAN is virtualized and hosted on open hardware with intelligence and machine learning capabilities. The O-DU, O-CU, and the near-RT RIC are considered VNFs that can be hosted on commercial off-the-shelf (COTS) servers [6]. Those O-RAN VNFs interact with each other forming a SFC where near RT-RIC communicates with O-CU (dependency relation 1) and O-CU communicates with O-DU (dependency relation 2). Those SFCs should be optimally placed on regional or edge clouds while maintaining the O-RAN operational requirements.

**Table I.** BIP Parameters and Variables

| Parameter/variable | Description |
|---|---|
| $\mathbb{S} = \{1, ..., s\}$ | Set of all servers |
| $\mathbb{V} = \{1, ..., v\}$ | Set of O-RAN units |
| $\mathbb{R}_v = \{1, ..., k\}$ | Set of redundant units |
| $\mathbb{D}_v = \{1, ..., q\}$ | Set of dependant units |
| $\mathbb{S}_{Regional} = \{1, ..., s_r\}$ | Set of servers located in the regional cloud |
| $\mathbb{S}_{Edge} = \{1, ..., s_e\}$ | Set of servers located in the edge cloud |
| $Dep$ | Dependent unit |
| $Rd$ | Redundant unit |
| $\lambda_i$ | Failure rate of node $i$ (unit or server) |
| $\delta_{ij}$ | Latency between two VNF units $i$ and $j$ |
| $\delta_{ss'}^{Server}$ | Latency between two servers $s$ and $s'$ |
| $\delta_{ij}^{Threshold}$ | Allowed latency between two VNF units $i$ and $j$ |
| $\delta_{ij}^{Rd}$ | Latency between a unit $i$ and its redundant $j$ |
| $\theta_{is}$ | Placement decision variable of unit $i$ on server $s$ |
| $\tau_i^R$ | Resources requiremnt for unit $i$ where $R$ can be CPU (C) or Memory (M) |
| $\tau_s^{R_s}$ | Available resources of server $s$ where $R_s$ can be CPU (C) or Memory (M) |
| $\beta_i^P$ | Binary value indicating if unit $i$ should be placed on the edge or regional cloud |
| $t_i^R$ | MTTR of node $i$ |

When an O-RAN is deployed using one of the scenarios defined by the O-RAN Alliance in [7], the logical network functions (near-RT RIC, O-CU, O-DU) are hosted as VNFs on the O-Cloud. As no system is perfect, software- or hardware-related faults and failures can occur and jeopardize the entire network operation by affecting service to the end-user. In Self Organizing Networks (SONs), healing from such events must be handled autonomously and rapidly since the outage of VNFs can disrupt services and have a catastrophic effect on mission-critical-applications. This section discusses the availability modeling and the proposed approach to mitigate O-RAN-related outages.

*A. Availability modeling*

The O-RAN alliance outlines all the deployment scenarios for the O-RAN VNFs that can be hosted on three locations: regional cloud, edge cloud, and cell site. However, the task of placing these VNFs is challenging and critical to the network's performance, health, recovery, and failure tolerance due to the various physical conditions at different cloud locations. For instance, regional clouds can have huge datacenters with high-performance servers while edge cloud servers have limited resources and performance. Choosing a deployment on highly reliable servers can greatly reduce the network's resiliency, enhance its self-healing abilities and its overall availability

With this in mind, we propose a BIP optimization model for a downtime-aware deployment strategy for the O-RAN VNFs. The optimization model considers the operational and performance constraints of the O-RAN with the goal of not only minimizing the per-VNF downtime, but also the SFC downtime in case of a failure or fault in the VNF or its hosting server. This placement strategy deploys the requested VNFs and their redundant instances on servers with high Mean Time to Failure (MTTF) and low Mean Time to Repair (MTTR) values to ensure the probability of a failure is minimized. It is necessary to note that each node (VNF or host) has its own MTTF, MTTR, and Recovery Time (RT). The availability calculations of O-RAN deployment depend on these three operational measures. They are defined as follows:

- MTTF: This metric defines the lifespan of a node before it stops operating. The MTTF is inversely proportional to the failure rate ($\lambda$).
- MTTR: This metric represents the average time needed to repair a node upon its failure.
- Recovery Time: This metric represents the failover time of a node's workload to its redundant ones if any.

The downtime reflects the time when a VNF fails and becomes unavailable to the network traffic until it is repaired or recovered where the traffic is re-routed to its redundant one. That said, the optimization model considers the placement of all redundant units in the network to ensure that when a failure occurs, those units are hosted in servers that allow the network to operate without violating any operational and performance constraints such as latency and dependency requirements. The availability of each node ($\alpha_{node}$) is calculated as follows.

$$\alpha_{node} = \frac{MTTF_{node}}{MTTF_{node} + MTTR_{node}} \quad (1)$$

We design our proposed solution using the scenario B deployment use case defined by the O-RAN Alliance because it is the primary focus for the assessment of O-RAN cloudification and its support to latency-aware/sensitive applications [7]. Fig. 2 depicts such a scenario where the near-RT RIC type VNFs should be hosted on the regional cloud and the O-CU and O-DU type VNFs should reside on the edge cloud.

*B. Mathematical formulation*

We propose a BIP model to solve the above-mentioned challenges and serve as a benchmark for downtime-aware VNF deployments in the O-RAN. This section covers the notation, objective function, and the constraints of the model.

*1) Notations and decision variables*

This section defines the decision variables presented in equation (2) and parameters of the BIP model. Table I. lists all the variables and parameters of the BIP model with their descriptions.

$$\theta_{is} = \begin{cases} 1 & if\ i\ is\ hosted\ on\ s \\ 0 & Otherwise \end{cases} \quad (2)$$

$$\theta_{is} \in \{0,1\} \quad \forall i \in \mathbb{V}, \forall s \in \mathbb{S}$$

$$\delta_{ij} \geq 0 \ \forall i, j \in \mathbb{V}$$

*2) Objective Function*

Our objective is to maximize availability thus minimizing the downtime of the deployed O-RAN VNFs. The availability of a single VNF component is calculated as shown in equation (1). The network is considered available when all its components (VNFs and servers) are available; thus, it is a series configuration. Said that, the network availability depends on the VNFs' and servers' failure rates (failure length and incidents), and the repair time [12]. Therefore, when selecting the host for a corresponding VNF, the model uses equations (3) (4) to filter out servers that maximize the new MTTF of the deployed VNF (equation 3) and minimize its new MTTR (equation 4). In this case, the failure rate and repair time of the VNF when it is hosted, are affected by its own MTTF/MTTR and those of its host as shown in the following equations.

$$MTTF_{VNF}^{Hosted} = \frac{1}{\lambda_{VNF} + \lambda_{Server}} \quad (3)$$

$$MTTR_{VNF}^{Hosted} = MTTR_{VNF} + MTTR_{Server} \quad (4)$$

Combining equations (3) and (4) into the availability calculation, we formulate the objective function (5) for the VNFs set $\mathbb{V}$ and their corresponding servers set $\mathbb{S}$ as follows. This objective function aims at maximizing the availability of the whole network and its units.

$$\max \sum_i^{|\mathbb{V}|} \sum_s^{|\mathbb{S}|} \left( \left( \frac{\frac{1}{\lambda_i + \lambda_s}}{\frac{1}{\lambda_i + \lambda_s} + (t_i^R + t_s^R)} \right) \times \theta_{is} \right) \quad (5)$$

*3. VNF Latency Constraints*

To minimize the downtime across the deployed VNFs, the proposed model ensures that the VNFs and their generated SFCs operate properly to maintain QoS and meet the Service Level Agreements (SLAs). The latency between the VNFs components must not exceed its operation threshold in the O-RAN architecture for O-Cloud deployments [7]. The optimization model ensures that any two communicating O-RAN units (whether redundant or dependent units) are hosted on servers that satisfy their delay requirements. For instance, given scenario B from the O-RAN-defined deployment scenarios, the maximum one-way delay (OWD) between the near-RT RIC type VNF and O-CU type VNF is 1ms [7]. In this case, the near-RT RIC should be placed in the regional cloud, and all O-CU/DU should be placed in the edge cloud while maintaining its delay constraint. The following equations reflect the latency constraints between the O-RAN units. Constraints (6) and (8) ensure that the latency between two O-RAN intercommunicating units $i$, $j$ (dependent or redundant units) are hosted on server(s) $s$, $s'$ that satisfy their latency requirements/threshold. Constraints (7) and (9) ensure that the latency between the redundant O-RAN units and its intercommunicating ones $i$, $j$ (dependent or redundant units) are hosted on server(s) $s$, $s'$ that satisfy their latency requirements/threshold. These constraints ensure that upon failure of the active unit, its redundant can handle the workload and communicate with other units without affecting the network performance and SLA. They also ensure that different O-RAN units can communicate properly without any service degradation.

$$\delta_{ss'}^{server} \times (\theta_{is} + \theta_{js'} - 1) - \delta_{ij}^{rd} < 0 \quad (6)$$

$$\forall i, j \in \mathbb{V}, \ \forall s, s' \in \mathbb{S}$$

$$\delta_{ss'}^{server} \times (\theta_{is} + \theta_{js'} - 1) - \delta_{ij}^{rd} < 0 \quad (7)$$

$$\forall i \in \mathbb{V}, \forall j \in \mathbb{R}_v, \forall s, s' \in \mathbb{S}$$

$$\delta_{ij} < \delta_{ij}^{Threshold} \quad \forall i, j \in \mathbb{V} \quad (8)$$

$$\delta_{ij} - \delta_{ij}^{Rd} \geq 0 \quad \forall i, j \in \mathbb{V} \quad (9)$$

*4. Computational Resources Constraints*

The following constraints ensure that a server $s$ has enough resources in terms of CPU cores *(C)* and memory *(M)* when hosting one or more VNFs $i$.

$$\sum_i^{|\mathbb{V}|} \theta_{is} \times \tau_i^c \leq \tau_s^{c_s} \quad \forall s \in \mathbb{S} \quad (10)$$

$$\sum_i^{|\mathbb{V}|} \theta_{is} \times \tau_i^M \leq \tau_s^{M_s} \quad \forall s \in \mathbb{S} \quad (11)$$

*5. Regional and Edge Cloud Constraints*

Two server sets are considered for the deployment of the VNFs in scenario B. Set one is located in the regional cloud, and the proposed model provides optimized placements of the near-RT RIC VNFs on those servers. On the other hand, set two is located in the edge cloud to host the O-CU and O-DU instances. Constraints (12) and (13) ensure that near-RT RIC type VNFs can only be hosted in the regional cloud while O-CU or O-DU type VNFs can be only deployed in the edge.

$$\sum_{s_r=0}^{\mathbb{S}_{Regional}} (\theta_{is_r} \times \beta_i^{Regional}) = 1 \quad \forall i \in \mathbb{V} \quad (12)$$

$$\sum_{s_e=0}^{\mathbb{S}_{Edge}} (\theta_{is_e} \times \beta_i^{Edge}) = 1 \quad \forall i \in \mathbb{V} \quad (13)$$

*6. Co-location & Anti-location Constraints*

Anti-location constraint (14) is designed to ensure that the principal VNF $i$ and its redundant $k$ should be placed on two different servers. Such a constraint maximizes the continuity of the units' operation. This constraint is also used for a unit *i* and its dependants *k* if the latter can operate in the absence of its sponsor. If the dependant unit cannot operate in the absence of its sponsor upon a failure, co-location constraint (15) is enabled. In this constraint, both the dependant and its sponsor can share the same node to maintain the operation's continuity and minimize the VNFs' downtime.

$$\theta_{is} + \theta_{ks} \leq 1 \quad (14)$$
$$\forall i \in \mathbb{V}, \ \forall k \in \mathbb{R}_v \cup \mathbb{D}_v, \ \forall s \in \mathbb{S}$$

$$\theta_{is} + \theta_{ks} \leq 2 \quad (15)$$
$$\forall i \in \mathbb{V}, \ \forall k \in \mathbb{D}_v, \ \forall s \in \mathbb{S}$$

## IV. EVALUATION AND USE-CASE

To test and validate the proposed optimization model, we have designed a naïve first-fit-first (NF3) greedy algorithm that represents the existing deployment approaches, which tackle the challenges of O-RAN VNF placement to maintain its performance. However, to demonstrate the advantages of our BIP model in terms of availability aspects, NF3 focuses on computational and operational latency constraints to ensure that the proposed deployment is valid. NF3 overlooks optimizing the placement of the redundant components on servers with high MTTF and low MTTR values, which does not allow the self-healing policy to optimally recover to those instances. A detailed explanation of the NF3 algorithm is presented in the following section.

### A. Naïve First-Fit-First (NF3) Algorithm

The NF3 algorithm accepts three inputs: the regional and edge cloud infrastructures, as well as the O-RAN VNF components and their redundancies. The algorithm starts by splitting the requested VNFs into three sets depending on their type (near-RT RIC, O-CU, O-DU). After the split, the algorithm searches the regional cloud server set to host the requested near-RT RIC type VNFs. The algorithm places the VNFs on the server with enough resources. Once the near-RT RIC type VNFs are placed, the algorithm searches the edge cloud server set to host the O-CU VNFs. Once a server with enough resources is found, the placement decision is executed based on the *eval()* function. The latter determines if the candidate server meets the latency constraints between the O-CU and the deployed near-RT RIC VNFs. The same procedure follows for the O-DU VNFs. However, the latency constraint at this stage is evaluated with the placed O-CU VNFs. Once the deployments for the VNFs are generated, the NF3 calculate their corresponding downtime and availability. It is important to highlight that in contrast to our proposed optimization model, NF3 overlooks availability constraints when selecting the candidate servers.

Due to the BIP model's time complexity, we test our proposed solution on a small-scale dataset of 50 servers distributed on the regional and edge clouds and a total of 21 VNFs (principal and redundant). The NF3 places all requested VNFs *(V)* by searching all candidate servers *(S)* for the server that meets the resource (CPU, memory) and location (regional, edge) constraints for each VNF. Since our proposed approach considers

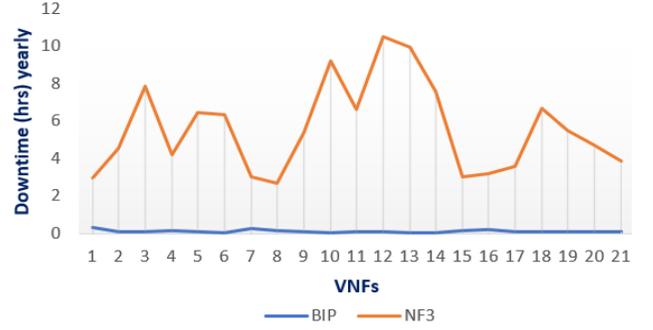

**Fig. 3.** The yearly per-VNF downtime comparison between the proposed BIP model and the NF3 greedy algorithm

3 types of VNFs (near-RT RIC, O-CU, O-DU), the server infrastructure is searched three times for each VNF type. This results in a complexity of *O(V× S)* for the greedy NF3 algorithm.

### B. Simulation Environment and Evaluation Metrics

Each candidate server has its computational parameters for in terms of CPU and memory, MTTF, MTTR, RT, and link delays to all servers within the cloud infrastructure. MTTF follows an exponential distribution with a mean of 3500 hours while MTTR and RT follow a normal distribution with means 2 and 0.5 hours, and a standard deviation of 1.5 and 0.016 respectively [13] [14] [15]. Candidate servers are divided into regional cloud servers for hosting near-RT RIC VNFs and edge cloud servers for hosting O-CU and O-DU VNFs. Every VNF has its computational requirements in the form of CPU and memory, its availability measures (MTTF, MTTR, RT). MTTF follows an exponential distribution with a mean of 2100 hours while MTTR and RT follow a normal distribution with means of 0.05 and 0.008 hours and standard deviation of 0.03 and 0.005 hours respectively [13] [14] [15]. In addition, the VNF set includes the redundancies for each VNF and the dependencies for every VNF (O-DU has a direct one to one dependency with the O-CU that has a direct many to one dependency with the near-RT RIC). The BIP model and the NF3 algorithm generate the deployment of each VNF and calculate its downtime and availability per year (per 8765 hours) accordingly. The downtime of the deployed VNF is affected by the failure rate and recovery time of itself and its host, and it is calculated in equations (16)(17)(18).

This scheduling problem, formulated using a linear programming model, is proven to be NP-hard [16]. Therefore, the approach is evaluated for small networks. We run our BIP model using IBM CPLEX on an Intel 9[th] Gen I7-9750H 2.6 GHz CPU computing server with 16GB RAM. The NF3 algorithm is written using Python 3.8 and is executed on the same server.

$$downtime_{node} = \lambda_{node} \times RT_{node} \quad (16)$$

$$downtime_{VNF} = \left( \frac{RT_{VNF}}{MTTF_{VNF}} + \frac{RT_{server}}{MTTF_{server}} \right) \quad (17)$$

$$\alpha_{VNF} = \frac{8765 - downtime_{VNF}}{8765} \quad (18)$$

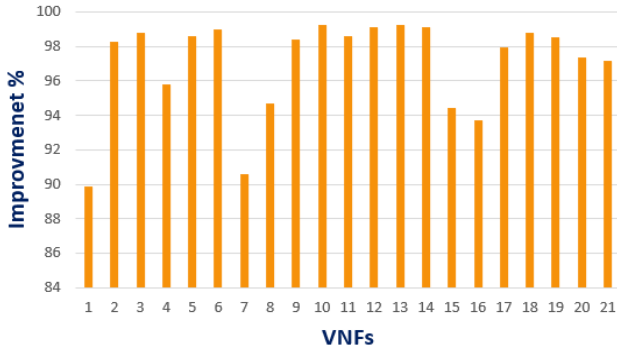

**Fig. 4**. The downtime improvement percentage of the BIP model over NF3

## V. Results

The proposed BIP optimization model is compared with the NF3 algorithm to check for the optimal placement strategy for the O-RAN while ensuring the minimum downtime across the network. The proposed BIP model can deploy the requested VNFs (principal and redundant) on servers in both the regional and edge clouds while adhering to all constraints. We measure the yearly availability of each VNF to demonstrate that not only is the overall network availability maximized, but the per-VNF availability is improved as well. The BIP model guarantees an average of %99.998 availability across all network VNFs while the NF3 can only achieve %99.935 average per VNF availability across the network. The NF3 approach selects the servers based on the performance requirements (latency and resources) while overlooking the availability constraints in terms of high operational metrics. It also discards the impact of the intercommunication relationship between units (redundancy or dependency) on their locations. The impact of these constraints over one year is clearly highlighted in the downtime. Fig. 3 shows the downtime reduction that the BIP achieves over NF3. The BIP model generates an average of 0.125 hours of downtime per year across all VNFs, while the NF3 results in 5.620 hours of downtime per year. The downtime improvement (in %) of the BIP model over the NF3 algorithm is shown in Fig 4. The BIP model achieves an average of 97.015% improvement in the experienced downtime over NF3 across all VNFs.

Although the proposed BIP model is proved to enhance the network's self-healing capabilities for O-RAN, its complexity hinders its applicability to large scale networks. That said, we aim to develop a heuristic solution and extend this work to large scale deployment scenarios for the O-RAN in the cloud.

## VI. Conclusion

The O-RAN architecture is introduced by the O-RAN Alliance to revolutionize the RAN by providing openness and intelligence to mobile networks. As the O-RAN is designed with intelligence, it is considered as a self-organizing network where self-healing is a key feature in intelligently handling and managing failures and faults in the network. To optimize its self-healing, we proposed a BIP model to optimize the placement of the requested VNFs and their redundant ones with the goal of maximizing the availability. The latter was achieved by employing a placement to minimize the per-VNF and the SFCs' downtime. The BIP model significantly outperforms the designed NF3 algorithm with an average of 97.015% yearly downtime improvement. Thus, such a model is considered the first downtime-aware building block toward providing a reliable solution for self-healing in O-RAN. In future work, the complexity of the proposed approach will be mitigated using machine learning models for the purpose of outage management/compensation for a self-healing O-RAN architecture.

## VII. Acknowledgement

This work is supported in part by Ciena Canada and Ontario Centre of Innovation (OCI).